\documentclass[manuscript]{acmart}
\usepackage{braket}
\usepackage{quantikz}
\usepackage{multirow}
\usepackage{graphicx}
\usepackage{caption}
\usepackage{subcaption}

\usepackage{booktabs}
\usepackage{color, colortbl}
\definecolor{LightGrey}{rgb}{0.88,0.88,1}
\definecolor{LightGreen}{rgb}{0.88,1,0.88}
\definecolor{LightRed}{rgb}{1,0.88,0.88}
\usepackage{adjustbox}

\begin{document}

\title{QuL: Programming Library for Computational Cooling of Qubits}

\author{Giuliano Difranco}
\affiliation{%
  \institution{University of Pisa}
  \city{Pisa}
  \country{Italy}
}
\email{g.difranco@studenti.unipi.it}
\orcid{0009-0008-6043-6048}

\author{Lindsay Bassman Oftelie}
\orcid{0000-0003-3542-1553}
\affiliation{%
  \institution{NEST, Istituto Nanoscienze-CNR and Scuola Normale Superiore}
  \city{Pisa}
  \country{Italy}
}
\email{lindsay.oftelie@nano.cnr.it}


\begin{abstract}
A key hurdle to the success of quantum computers is the ability to initialize qubits into a pure state, which can be achieved by cooling qubits down to very low temperatures. Computational cooling of qubits, whereby a subset of the qubits is cooled at the expense of heating the other qubits via the application of special sets of logic gates, offers a route to effectively cool qubits.  Here, we present QuL, a programming library which can be used to generate, analyze, and test quantum circuits for various computational cooling protocols.  In its most basic usage, QuL enables a novice user to easily produce cooling circuits with minimal input or knowledge required.  The programming library, however, offers flexibility to more advanced users to finely tune the cooling protocol used to generate the quantum circuit.  Finally, QuL offers methods to assess and compare various cooling protocols for users interested in studying optimal implementation of computational cooling in general, or on specific quantum backends. It is our hope that QuL will not only facilitate the execution of computational cooling on current quantum computers, but also serve as a tool to investigate open questions in the optimal implementation of computational cooling.
\end{abstract}

\maketitle

\section{Introduction}
Quantum computers are on the cusp of revolutionizing scientific computing.  Capitalizing on quantum effects, such as superposition and entanglement, quantum computers offer exponential advantages over classical computers for certain problems, including optimization, cryptography, and simulation of quantum systems \cite{preskill1998reliable, bassman2021simulating}.  Various prototypes of quantum computers are actively under development, and a consensus for the optimal implementation, in terms of performance and scalability, has not yet been reached. 

Regardless of the specific implementation, however, all quantum computers must satisfy a set of requirements known as DiVincenzo's criteria \cite{divincenzo2000physical}.  One of these requirements is the ability to initialize the qubits into a known, pure quantum state.  This requirement stems from the elementary computing requirement (quantum or classical) that memory registers must be initialized in a known, fiducial state before beginning computation.  Furthermore, fault-tolerant quantum computers of the future will require ancillary qubits initialized into pure states throughout the course of computation to perform quantum error correction \cite{knill1998resilient, aharonov1997fault,preskill1998reliable}.  A key hurdle to the success of quantum computers, therefore, is the ability to initialize qubits into a very pure (i.e., cold) state.  

Various methods exist to directly cool qubits via physical mechanisms, such as with magnetic fields or lasers.  However, such techniques are generally resource intensive and have limitations due to physical feasibility.  Scientists, therefore, began to explore means of effectively cooling qubits via the action of special sets of logic gates, which we refer to as \textit{computational cooling}.  The last few decades have seen the development of a number of computational cooling techniques, including dynamic cooling \cite{schulman1999molecular, bassman2024dynamic} and heat-bath algorithmic cooling \cite{rempp2007cyclic, kaye2007cooling, elias2011semioptimal, rodriguez2017correlation, raeisi2019novel}.

Here, we present QuL, an open-source programming library which can be used to generate, analyze, and test quantum circuits used to execute various computational cooling protocols.  With minimal background knowledge or experience, a novice user can automatically generate a cooling circuit by solely entering the desired number of qubits to be utilized by selecting one of the pre-defined cooling protocols included in QuL.  The programming library, however, offers flexibility for advanced users to finely tune built-in cooling protocols or even develop their own cooling protocols.  QuL also offers methods to assess and compare various cooling protocols for users interested in studying optimal implementations of computational cooling both in general and on specific quantum backends.  We provide two illustrative examples to demonstrate important use cases for QuL. The full code, as well as a tutorial can be found on GitHub \cite{github}. By simplifying the preparation and analysis of computational cooling circuits, QuL provides a direct path for researchers seeking to implement computational cooling on near-term quantum computers, as well as a toolkit for investigating optimal implementations of computational cooling in the future.

\section{Computational Cooling Methods}\label{sec:methods}
\subsection{Dynamic Cooling}
\begin{figure}[h]
  \centering
  \hspace{5mm}
  \begin{subfigure}[c]{0.45\textwidth}
        \includegraphics[width=\textwidth]{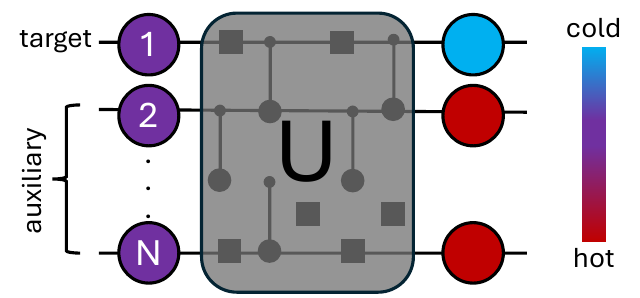} 
        \put(-200,90){(a)}
        \put(15,90){(b)}
    \end{subfigure}
  \hspace{8mm}  
  \begin{subfigure}[c]{0.45\textwidth}
        \begin{quantikz}[row sep=5mm,column sep=5mm]
        \lstick{$q_1$} & \ctrl[open]{1} & \ctrl[open]{1}& \targ{0} & \ctrl[open]{1}&\ctrl[open]{1} & \rstick{} \\
        \lstick{$q_2$} & \ctrl{1} & \targ{1} & \ctrl[open]{-1} & \targ{1} & \ctrl{1} & \rstick{} \\
        \lstick{$q_3$} & \targ{2} & \ctrl[open]{-1} & \ctrl[open]{-1} & \ctrl[open]{-1} & \targ{2} & \rstick{} 
        \end{quantikz}
    \end{subfigure}
  \caption{Dynamic cooling of qubits.  (a) Schematic of dynamic cooling with $N$ identical qubits.  Here, the target qubit is cooled at the expense of heating up the auxiliary qubits via the application of a global unitary operator $U$, which is carried out via a quantum circuit. (b) A quantum circuit implementing dynamic cooling with 3 qubits.  Here, $q_1$ is the target qubit to be cooled and $q_2$ and $q_3$ are auxiliary qubits. }
  \Description{Schematic of a quantum circuit for dynamic cooling.}
  \label{fig:schematic}
\end{figure}

The first method for computational cooling of qubits was proposed in 1999 \cite{schulman1999molecular}. Known as \textit{dynamic cooling}, the method is based on entropy manipulation in a closed system. Specifically, a set of $N$ qubits is assumed to be initialized at an initial temperature $T$.  A global unitary (i.e., reversible) operator $U$ is then enacted on the total system of qubits via a quantum circuit.  The effect is that a single target qubit is cooled at the expense of heating up the other $N-1$ auxiliary qubits, as shown schematically in Figure \ref{fig:schematic}a. The method was all but abandoned shortly after its inception due to its poor scaling of cooling with system size $N$ in the high initial temperature regime \cite{boykin2002algorithmic, fernandez2004algorithmic}. The nuclear magnetic resonance (NMR) quantum computers available at the time were comprised of qubits initialized at high effective temperatures, and it was determined that an unreasonably large system size (on the order of $N \approx 10^6$) would be needed to sufficiently cool the target qubit.  

Modern implementations of quantum computers, however, have shifted away from NMR devices. Contemporary quantum computers, comprising qubits made from superconducting circuits, trapped ions, and neutral atoms, all now operate at extremely low initial temperatures.  It was recently discovered that dynamic cooling exhibits superior scaling with system size when the qubits are initialized at low temperatures \cite{bassman2024dynamic}.  Specifically, there exists a crossover, whereby the final temperature after cooling scales as $T/\sqrt N$ in the high initial temperature regime, but scales as $T/N$ in the low initial temperature regime.  This marked improvement in scaling reinstates the interest in dynamic cooling for current low-temperature quantum computers.  

While the maximal amount of cooling for a given total system size $N$ via this method has been determined \cite{bassman2024dynamic}, the optimal implementation of dynamic cooling, in terms of quantum circuits, remains an open question.  Indeed, there exists a large family of unitaries $U$ that can enact maximal cooling, and each unitary will carry a different level of complexity in terms of the size of the quantum circuit it generates.  Ref. \cite{bassman2024dynamic} discusses various advantages and disadvantages of three protocols for producing different maximally cooling unitaries and proposes a metric for determining the complexity of the associated quantum circuits.  Figure \ref{fig:schematic}b shows one possible quantum circuit implementation of dynamic cooling with $N=3$ qubits, though other quantum circuits could be derived which achieve equivalent cooling of the target qubit.  Here, the qubit $q_1$ is cooled at the expense of heating up qubits $q_2$ and $q_3$. This quantum circuit is comprised of 3-qubit multi-controlled-NOT gates, known as Toffoli gates, which apply a NOT gate (represented with a circle with a cross) to the associated qubit dependending on the state of the other two control qubits.  Closed circles in the Toffoli gates implies the NOT gate is applied when the corresponding control qubit is in the $\ket{1}$ state, while open circles imply the NOT gate is applied when the corresponding control qubit is in the $\ket{0}$ state.

QuL provides functions to implement dynamic cooling according to three different protocols, which include the Partner-Pairing Algorithm (PPA) \cite{schulman2005physical}, the minimum work protocol, and the mirror protocol (see Appendix A of Ref. \cite{bassman2024dynamic} for their definition and more details).  While all three protocols will result in maximal cooling of the target qubit, the best protocol to use will depend on the specifics of the use case.  The user is also free to provide their own maximally cooling unitary (according to some user-defined protocol) to implement dynamic cooling within QuL.    

\subsection{Sub-Optimal Dynamic Cooling}
The greatest obstacle to the success of dynamic cooling on near-term quantum computers is the complexity of the cooling circuit. Indeed, the complexity of the quantum circuit grows quickly with the total number of qubits $N$ used for cooling. One way to mitigate this problem is to use a sub-optimal dynamic cooling routine, whereby instead of cooling a target qubit to the optimal (i.e., minimum) final temperature, we agree to a slightly reduced the amount of cooling in order to achieve a massive reduction in the quantum circuit complexity.

QuL provides functions to generate quantum circuits for one such suboptimal cooling algorithm, derived and characterized in Ref. \cite{bassman2024dynamic}, which is defined as follows. Consider a system with a total of $N=n^r$ qubits, divided into $N/n = n^{r-1}$ clusters each containing $n$ qubits.  Cooling is executed in $r$ steps.  In the first step, dynamic cooling is performed within each of the clusters, such that one target qubit in each of the clusters is maximally cooled for a system size of $n$ (we emphasize that this will necessarily be less cooling than the maximal cooling allowed for a system of size $N$). In the next step, the $n^{r-1}$ cooled qubits are again divided into clusters of size $n$.  There will now be $n^{r-2}$ of these clusters, and dynamic cooling is once again applied within each cluster, resulting in $n^{r-2}$ further cooled qubits.  This is repeated until $r$ such rounds of cooling have been executed, resulting in one ultimately cooled target qubit.  The final temperature of this target qubit is greater than the minimum final temperature that could have been achieved with dynamic cooling on $N$ qubits.  However, the complexity of the quantum circuit for this suboptimal cooling scheme is drastically reduced compared to the circuit complexity for dynamic cooling on $N$ qubits.  This is because in the suboptimal cooling routine, the total quantum circuit is only comprised of a finite number of subcircuits each acting on the subspace of $n$ qubits, as opposed to $N$ qubits.  While dynamic cooling requires a circuit to be generated from a unitary matrix of size $2^N \times 2^N$, the suboptimal cooling circuit is generated from unitary matrices of size $2^n \times 2^n$. Since $n=\sqrt[^r]{N}$, the massive reduction in the size of the associated unitary matrices directly translates to a dramatic reduction in the complexity of the quantum circuit.

\begin{figure}[h]
  \centering
  \hspace{5mm}
  \begin{subfigure}[c]{0.25\textwidth}
        \includegraphics[width=\textwidth]{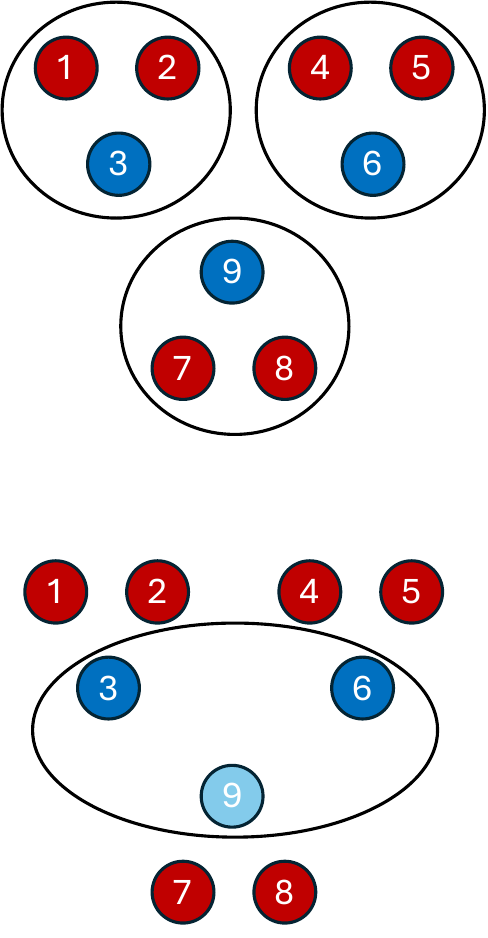} 
        \put(-130,230){(a)}
        \put(-100,210){$1^{st}$ Round of Cooling}
        \put(-100,90){$2^{nd}$ Round of Cooling}
        \put(10,230){(b)}
        \put(55,220){$1^{st}$ Round of Cooling}
        \put(150,220){$2^{nd}$ Round of Cooling}
    \end{subfigure}
  \hspace{8mm}  
  \begin{subfigure}[c]{0.45\textwidth}
        \begin{quantikz}[row sep=5mm,column sep=3mm]
        \lstick{$q_1$} & \targ{2} & \ctrl[open]{1} & \ctrl[open]{1} & \ctrl[open]{1} & \targ{2} & \slice{} \rstick{} \\
        \lstick{$q_2$} & \ctrl{-1} & \targ{1} & \ctrl[open]{1} & \targ{1} & \ctrl{-1} & \rstick{} \\
        \lstick{$q_3$} & \ctrl[open]{-1} & \ctrl[open]{-1}& \targ{0} & \ctrl[open]{-1}&\ctrl[open]{-1} & &\targ{2} &  \ctrl[open]{3} & \ctrl[open]{3} & \ctrl[open]{3} & \targ{3} & \rstick{} \\
        \lstick{$q_4$} & \targ{2} & \ctrl[open]{1} & \ctrl[open]{1} & \ctrl[open]{1} & \targ{2} &  \rstick{} \\
        \lstick{$q_5$} & \ctrl{-1} & \targ{1} & \ctrl[open]{1} & \targ{1} & \ctrl{-1} &  \rstick{} \\
        \lstick{$q_6$} & \ctrl[open]{-1} & \ctrl[open]{-1}& \targ{0} & \ctrl[open]{-1}&\ctrl[open]{-1} && \ctrl{-3} & \targ{1} & \ctrl[open]{3} & \targ{1} & \ctrl{-3} &  \rstick{} \\
        \lstick{$q_7$} & \targ{2} & \ctrl[open]{1} & \ctrl[open]{1} & \ctrl[open]{1} & \targ{2} &  \rstick{} \\
        \lstick{$q_8$} & \ctrl{-1} & \targ{1} & \ctrl[open]{1} & \targ{1} & \ctrl{-1} &  \rstick{} \\
        \lstick{$q_9$} & \ctrl[open]{-1} & \ctrl[open]{-1}& \targ{0} & \ctrl[open]{-1}&\ctrl[open]{-1} && \ctrl[open]{-3} & \ctrl[open]{-3}& \targ{0} & \ctrl[open]{-3}&\ctrl[open]{-3} & \rstick{}
        \end{quantikz}
    \end{subfigure}
  \caption{Suboptimal dynamic cooling of qubits using $r=2$ rounds of cooling with cluster sizes of $n=3$. (a) A schematic diagram of suboptimal dynamic cooling.  The large black circles represent the application of dynamic cooling to the enclosed qubits. (b) A quantum circuit executing suboptimal dynamic cooling.  The qubit labels of the wires in the circuit diagram correspond to the numbered qubits in panel (a).}
  \Description{Suboptimal dynamic cooling.}
  \label{fig:suboptimal_cooling}
\end{figure}
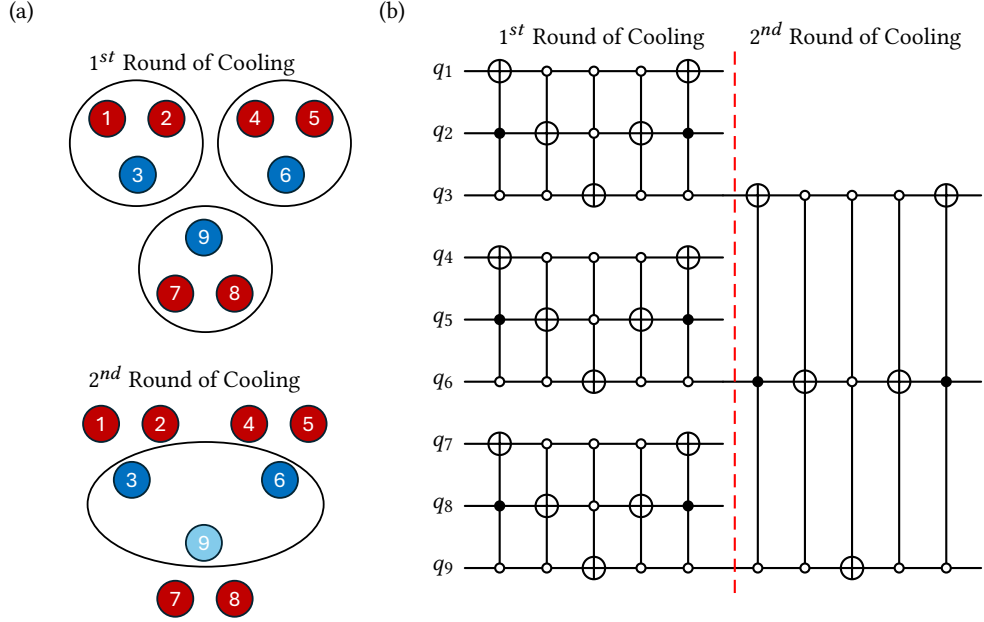

Figure \ref{fig:suboptimal_cooling}a shows a schematic and quantum circuit implementation of suboptimal dynamic cooling using $r=2$ rounds of cooling with clusters of size $n=3$.  A total of $N=9$ qubits are used to cool qubit $q_9$. The qubits are broken into three clusters of three qubits each.  In the first round of suboptimal dynamic cooling, qubits $q_3$, $q_6$, and $q_9$ are each cooled via dynamic cooling with three qubits.  In the second round, cooled qubits $q_3$, $q_6$, and $q_9$ undergo a second round of dynamic cooling, bringing qubit $q_9$ to an even cooler temperature.  Figure \ref{fig:suboptimal_cooling}b shows a quantum circuit implementing this suboptimal dynamic cooling routine.  Note that this circuit is only comprised of 3-qubit controlled-NOT gates, whereas dynamic cooling amongst $N=9$ qubits would have required 9-qubit controlled-NOT gates, which are significantly more complex to implement.

\subsection{Heat-Bath Algorithmic Cooling}
While dynamic cooling scales more advantageously in the low initial temperature regime, the fact that cooling occurs in a closed system (i.e., a system that does not interact with its environment) sets a lower bound on the final temperature that can be reached by the target qubit for a given system size $N$.  This limit is set by the so-called Shannon bound, which states that the total entropy in a closed system cannot decrease \cite{boykin2002algorithmic, fernandez2004algorithmic}.  Proposals for new computational cooling techniques, seeking to bypass this bound, therefore sought to extend the closed system to an open system by allowing a subset of qubits to interact with the environment (i.e., a heat bath), extending cooling capability below the Shannon bound.  In one such method, known as \textit{heat bath algorithmic cooling} (HBAC), a target qubit is cooled by alternating steps of (i) reversible entropy manipulation amongst the entire qubit set via unitary operations (i.e., dynamic cooling) and (ii) irreversible resetting of a subset of the heated qubits to the initial temperature.  This ability to beat the Shannon bound triggered a flurry of results in the field of HBAC: various implementations were proposed \cite{rempp2007cyclic, kaye2007cooling, elias2011semioptimal, rodriguez2017correlation, raeisi2019novel}; theoretical limits were explored \cite{schulman2005physical,ticozzi2014quantum, brassard2014prospects, raeisi2015asymptotic, rodriguez2016achievable}; and various experimental demonstrations were performed \cite{baugh2005experimental, fernandez2005paramagnetic, ryan2008spin, brassard2014experimental,  atia2016algorithmic} (see \cite{park2016heat} for a review).  

While HBAC can achieve lower temperatures than dynamic cooling, it presumes the availability of two different species of qubits: one to be used as computational qubits (i.e., the target qubit), and one to be used as reset qubits (which will interact with the heat bath to unload entropy from the system to the environment).  The computational qubits should have very slow relaxation times in order to maintain coherence throughout the computation, while the reset qubits should have very fast relaxation times in order to quickly re-thermalize with the bath at each iteration of cooling.  This assumption is reasonable within the setting of NMR quantum computers (which were the predominant implementation of quantum computers during the development of HBAC), where nuclear spins within the molecules can be selected for the computational qubits and electronic spins within the molecules can be selected for the reset qubits.  However, the majority of contemporary quantum computers now comprise identical qubits, usually optimized to have long relaxation times. This makes it difficult to implement HBAC in a striaghtforward way on current quantum computers.  It is possible that future quantum computing implementations could once again offer two species of qubits, so QuL therefore provides support for generating HBAC circuits by inserting "RESET" gates into the circuit, which imply that the qubit should be allowed to rethermalize with the bath.  While, these circuits can be used in theoretical studies of optimal implementation of HBAC, we emphasize that they should not be directly executed on current quantum computers with identical qubits, as qubits will typically not be able to be reset fast enough to successfully perform multiple cooling rounds.  However, it is possible to implement a variation on HBAC that can be executed with identical qubits, which we call \textit{semi-open algorithmic cooling} and describe in subsection \ref{sec:semi-open}.

\begin{figure}[h]
  \centering
  \hspace{5mm}
  \begin{subfigure}[c]{0.55\textwidth}
        \includegraphics[width=\textwidth]{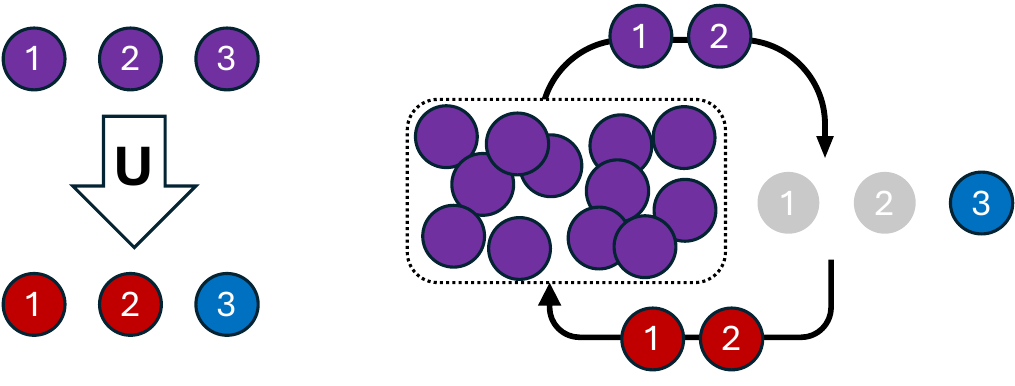} 
        \put(-250,90){(a)}
        \put(-250,-20){(b)}
    \end{subfigure}\\[2mm] 
  \begin{subfigure}[c]{0.5\textwidth}
    \begin{quantikz}[row sep=5mm,column sep=3mm]
\lstick{$q_1$} & \targ{2} & \ctrl[open]{1} & \ctrl[open]{1} & \ctrl[open]{1} & \targ{2} & \gate{RESET}  &  \targ{2} & \ctrl[open]{1} & \ctrl[open]{1} & \ctrl[open]{1} & \targ{2} & \rstick{} \\
\lstick{$q_2$} & \ctrl{-1} & \targ{1} & \ctrl[open]{1} & \targ{1} & \ctrl{-1} & \gate{RESET} &  \ctrl{-1} & \targ{1} & \ctrl[open]{1} & \targ{1} & \ctrl{-1} & \rstick{} \\
\lstick{$q_3$} & \ctrl[open]{-1} & \ctrl[open]{-1}& \targ{0} & \ctrl[open]{-1}&\ctrl[open]{-1} & &  \ctrl[open]{-1} & \ctrl[open]{-1}& \targ{0} & \ctrl[open]{-1}&\ctrl[open]{-1} & \rstick{}
\end{quantikz}    
    \end{subfigure}
  \caption{Heat-bath algorithmic cooling (HBAC) of $N=3$ qubits. (a) A schematic diagram of HBAC, which consists of repeated cycles of dynamic cooling and rethermalizing the reset qubits (here, qubits $q_1$ and $q_2$) with the bath. (b) A quantum circuit implementing two round of HBAC. The qubit labels of the wires in the circuit diagram correspond to the numbered qubits in panel (a).}
  \Description{HBAC}
  \label{fig:HBAC}
\end{figure}

Figure \ref{fig:HBAC} shows a schematic and quantum circuit implementation of HBAC with three qubits. In the first phase of HBAC (left panel of Figure \ref{fig:HBAC}a), qubit $q_3$ is cooled via dynamic cooling.  In the second phase (right panel of Figure \ref{fig:HBAC}a), the reset qubits $q_1$ and $q_2$ are reset to the bath temperature.  These two phases are alternately repeated until the target qubit $q_3$ reaches the desired final temperature. Figure \ref{fig:HBAC}b shows a quantum circuit implementing two rounds of HBAC. We again emphasize that the "RESET" gates shown in the circuit are not logical gates, but rather represent the corresponding qubit being reset to the initial temperature of the environment.

\subsection{Semi-Open Algorithmic Cooling}
\label{sec:semi-open}
Inspired by HBAC, semi-open algorithmic cooling (AC) is a computational cooling technique we introduce in this work, which can be executed on quantum computers consisting of identical qubits (unlike HBAC).  In its most straightforward application, cooling is carried out in multiple steps.  In the first step, a cluster of $n$ qubits all at an initial temperature $T$ is selected and dynamic cooling is applied to cool a target qubit.  This target qubit is then grouped with $n-1$ hitherto unutilized qubits in the quantum processor to form a new $n$-qubit cluster.  This time, howwever, notice that the qubits in the cluster are not all at the same initial temperature.  The target qubit is at the minimum final temperature achievable with dynamic cooling with $n$ qubits, while the auxiliary qubits are at the original initial temperature $T$.  A cooling unitary can then be applied to this cluster of qubits.  Since the initial temperatures are not all equal, the optimally cooling unitary may not be equivalent to a dynamic cooling unitary.  In this case, a new maximally cooling unitary may need to be derived.  Either way, the target qubit is further cooled, and this process can continue for arbitrarily many steps where $n-1$ previously unutilized qubits are selected in each round $r$ to group with the target qubit for application of a global cooling unitary.

\begin{figure}[h]
  \centering
  \begin{subfigure}[c]{0.45\textwidth}
        \includegraphics[width=\textwidth]{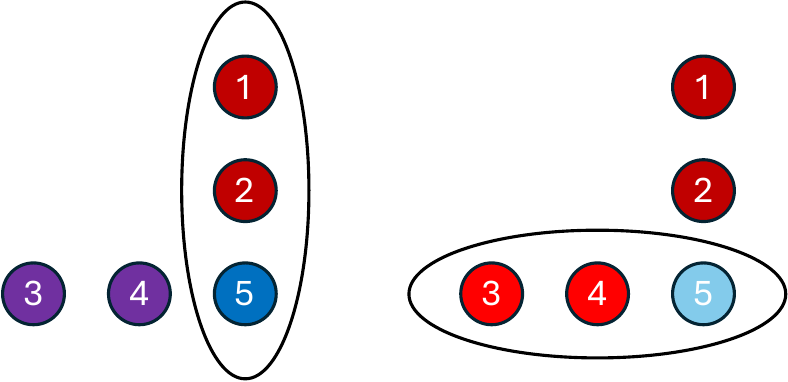} 
        \put(-200,100){(a)}
        \put(0,100){(b)}
    \end{subfigure}
      \hspace{5mm}
  \begin{subfigure}[c]{0.4\textwidth}
    \begin{quantikz}[row sep=5mm,column sep=3mm]
\lstick{$q_1$} & \targ{2} & \ctrl[open]{1} & \ctrl[open]{1} & \ctrl[open]{1} & \targ{2} \slice{} & \rstick{} \\
\lstick{$q_2$} & \ctrl{-1} & \targ{1} & \ctrl[open]{3} & \targ{1} & \ctrl{-1} & \rstick{} \\
\lstick{$q_3$}&&&&&&& \setwiretype{q} & \targ{2} & \ctrl[open]{1} & \ctrl[open]{1} & \ctrl[open]{1} & \targ{2} & \rstick{}\\
\lstick{$q_4$}&&&&&&& \setwiretype{q} & \ctrl{-1} & \targ{1} & \ctrl[open]{1} & \targ{1} & \ctrl{-1} & \rstick{}\\
\lstick{$q_5$} & \ctrl[open]{-3} & \ctrl[open]{-3}& \targ{0} & \ctrl[open]{-3}&\ctrl[open]{-3} &&& \ctrl[open]{-1} & \ctrl[open]{-1}& \targ{0} & \ctrl[open]{-1}&\ctrl[open]{-1} & \rstick{}
\end{quantikz} 
    \end{subfigure}
  \caption{Semi-open algorithmic cooling (AC) with $N=5$ qubits. (a) A schematic diagram depicting two rounds of semi-open AC. (b) A quantum circuit implementation executing semi-open AC. The qubit labels of the wires in the circuit diagram correspond to the numbered qubits in panel (a).}
  \Description{Semi-open algorithmic cooling}
  \label{fig:semiopenAC}
\end{figure}

Figure \ref{fig:semiopenAC}a shows a schematic and quantum circuit implementation of semi-open AC.  In the first round, qubits $q_1$ and $q_2$ are employed to cool qubit $q_5$ via dynamic cooling.  In the second round, qubits $q_3$ and $q_4$ are employed to further cool qubit $q_5$.  Notice, however, that in the second round of cooling, the qubits are not all initialized at the same temperature, meaning that the second round of cooling will differ from dynamic cooling (which assumes all qubits are initialized at the same temperature).  Figure \ref{fig:semiopenAC}b shows a quantum circuit implementation of this semi-open AC procedure.

It must also be noted that auxiliary qubits that will be used in later rounds of cooling must lie idle during earlier rounds of cooling.  As qubits have a finite decoherence time, it is important to limit the number of rounds of semi-open AC such that the total time for the cooling procedure does not become larger than the qubit decoherence times.  An error mitigation technique, known as dynamical decoupling (DD) \cite{viola1998dynamical, zanardi1999symmetrizing, vitali1999using, duan1999suppressing}, has been developed to suppress such decoherence error in idling qubits.  It works by applying a set of pulses (which together amount to application of the identity operator) to the idling qubits, which cancels the system-environment interaction \cite{pokharel2018demonstration}.  It may be prudent to apply DD to idling qubits during semi-open AC in order to achieve better results.

\section{Software Description}
\label{sec:software}
\begin{figure}
\centering
\includegraphics[scale=0.6]{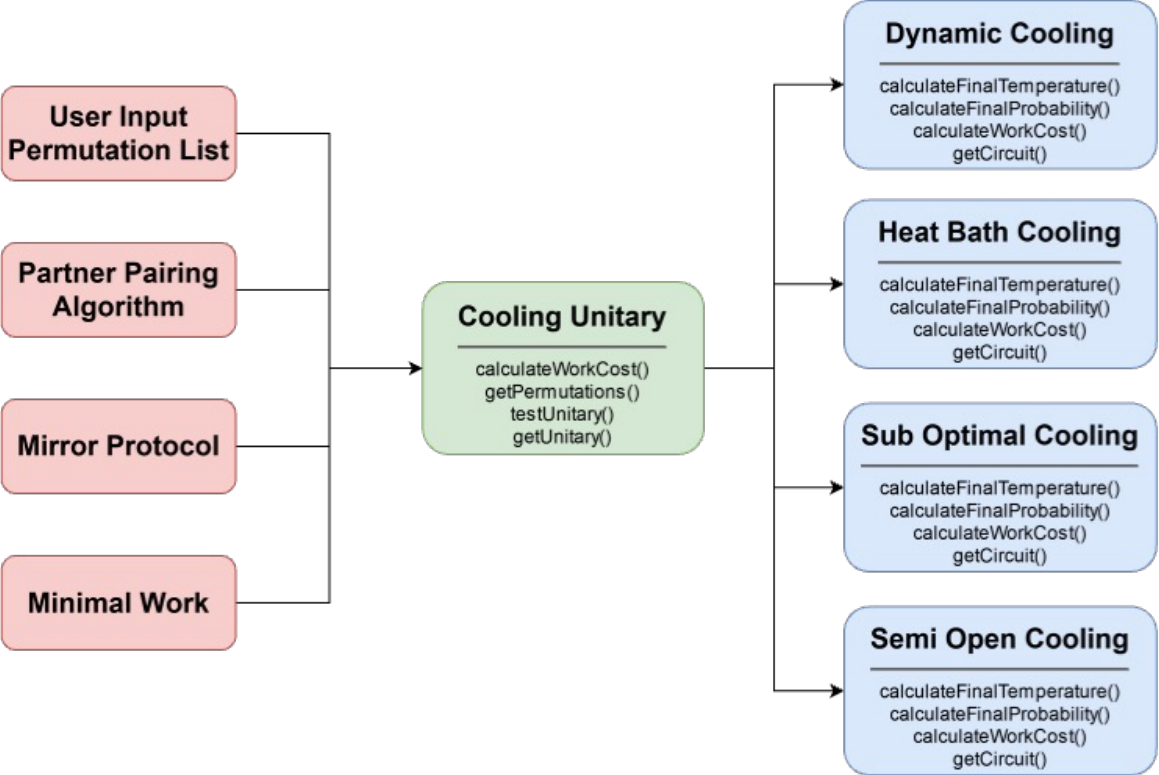}
\caption{Blueprint of the QuL programming library.  Blue rectangles in the right-most column denote the four main computational cooling modules currently implemented in QuL.  Each module has four main methods for analyzing various characteristics of the cooling procedure.  The primary data structure used to instantiate all computational cooling methods is the Cooling Unitary, shown in the green rectangle in the center column.  Required input to construct a cooling unitary can be generated by any of the three built-in protocols, shown in red rectangles in the left-most column.  Advanced users are also free to provide their own cooling protocol via a permutation list.}
\label{fig:blueprint}
\end{figure}

A blueprint for the QuL programming library is depicted in Figure \ref{fig:blueprint}. QuL currently supports four main techniques for computational cooling, namely dynamic cooling, sub-optimal dynamic cooling, heat-bath algorithmic cooling, and semi-open algorithmic cooling, which comprise the four main modules, depicted in blue in the rightmost column in Figure \ref{fig:blueprint}.  By organizing the code into a set classes each implementing a different computational cooling technique, QuL is an extensible package, where new modules can be easily added and integrated into the code as new computational cooling techniques are developed.  

All the computational cooling modules rely on the construction of a unitary operator (i.e., matrix), which can be applied to different subsets of qubits in order to enact cooling.  This cooling operator is stored within a primary data structure of QuL, called a Cooling Unitary, depicted as a green rectangle in the center of Figure \ref{fig:blueprint}.  In general, the cooling operator is a generalized permutation matrix, which can perform an arbitrary number of cyclic permutations of arbitrary lengths on the states of the subsystem on which it is applied. While a cyclic permutation of length two represents a SWAP between two states, cyclic permutations of length $m$ perform permutations of the form $\ket{i_1} \rightarrow \ket{i_2} \rightarrow ... \ket{i_m} \rightarrow \ket{i_1}$, where $\ket{i}$ represents a quantum state of the subsystem.  It is a generalize permutation matrix because non-zero entries in the matrix may contain arbitrary phases.

A Cooling Unitary in QuL, can therefore be instantiated by providing a list of permutations which defines the cooling operator, and the number of qubits on which this cooling operator will act.  This input can be provided with an explicit list of permutations, or an automatically generated list of permutations derived via built-in protocols, depicted in red in the leftmost column in Figure \ref{fig:blueprint}. If users have their own cooling protocol, they can build a Cooling Unitary by providing their own list of permutations. Each permutation in the list may be given as an ordered list of the states involved in the cyclic permutation.  The states can be specified in two forms: (i) a binary string with each character denoting whether the corresponding qubit is in the $0$ or $1$ state (e.g., the string $'101'$ denotes a state with the first and third qubits in the $1$ state and the second qubit in the $0$ state) or (ii) a decimal integer corresponding to the binary bitstring representation of the state (e.g., the integer 5 corresponds to the $101$ state).  The two forms can both be used within the same definition of a cyclic permutation.  For example, the cyclic permutation $\ket{000} \rightarrow \ket{001} \rightarrow \ket{100} \rightarrow \ket{000}$ can be specified as ["000","001","100"] or [0,1,"100"] (all permutations are assumed to be cyclic, and thus the initial state need not be listed at the end of the list).  

For ease of use, QuL also provides several built-in protocols which users can employ to automatically generate a Cooling Unitary by solely providing the number of qubits in the system.  These protocols include the partner-pairing algorithm (PPA), the mirror protocol, and the minimum work protocol.  These can be used with the classes PartnerPairingAlgorithm, MirrorProtocol, and MinimalWorkProtocol, respectively, which each return a Cooling Unitary derived from the respective cooling protocol. 

\begin{table}[ht]
\centering
\resizebox{11cm}{!}{%
    \begin{tabular}{clrlr}
    \toprule
    \midrule
    \multicolumn{1}{c}{}& \multicolumn{2}{c}{Explicit Array} & \multicolumn{2}{c}{CSR Array} \\ 
    \cmidrule(lr){2-3} \cmidrule(lr){4-5}
    N & Memory [bytes] & Time [seconds] &Memory [bytes] & Time [seconds] \\
    \toprule
    \rowcolor{LightGrey}
    8 & 262144  & $\sim$0.01 & 3076 & $\sim$0.001\\
    10 & 4194304  & $\sim$2.00 & 12292 & $\sim$0.001\\
    \rowcolor{LightGrey}
    12 & 67108864 & $\sim$790 & 49156 & $\sim$0.001 \\
    14 & 1073741824 & $\sim$72950 & 196612 & $\sim$0.001 \\
    \rowcolor{LightGrey}
    16 & 17179869184 & (*) & 786436 & $\sim$0.01 \\
    18 & 274877906944 & (*) & 3145732 & $\sim$0.02 \\
    \midrule
    \bottomrule
    \end{tabular}
    }
    \vspace{0.3cm}
\caption{Table describing the difference between a Cooling Matrix with NumPy and CSR. The first column describes the size of the Cooling Unitary, the second and the fourth column describes the space occupied by the Unitary in bytes for a given size. The third and the fifth column describes the time to calculate the dot product by two Unitaries with that size. 
\\(*) The Unitary is too large to fit main memory.}
\label{tab:CSRMAtrix}
\end{table}

The cooling operator is given by a matrix of size $2^N \times 2^N$, where $N$ is the number of qubits on which the operator will act.  Clearly, this grows exponentially with the size of the system, which could easily start to strain memory resources as the system size is increased.  Fortunately, however, permutation matrices are extremely sparce, which enables the Cooling Unitary class to efficiently store the cooling operator as a Compressed Sparse Row (CSR) array, a data structure provided by the SciPy python library \cite{2020SciPy-NMeth}, which supports fast matrix multiplication.  By using this optimized data structure, the Cooling Unitary class gains massive savings in both memory storage as well as compute time, as shown in Table \ref{tab:CSRMAtrix}.  The first column gives the number of qubits in the system $N$ on which the cooling operator acts.  The second and fourth columns show the amount of memory that is required to store the cooling operator in an explicit array and a CSR array, respectively.  While the memory requirement exponentially becomes intractable with increasing system size for an explicit array, the CSR array allows for efficient storage of the cooling operator.  The third and fifth columns show the amount of time it takes to perform matrix multiplication between two cooling operators (which is required by some protocols to generate the final cooling unitary) using explicit arrays and CSR arrays, respectively.  Once again, we see that the time requirement quickly becomes unreasonable when using explicit arrays, while using CSR arrays allows for significantly faster computation times.  While CSR arrays are clearly the more favorable way to store the cooling operator, the Cooling Unitary class does provide a method to convert the CSR array to an explicit array if required by the user.

Once a Cooling Unitary has been prepared, it can be passed to one of the computational cooling modules to generate a quantum circuit and analyze various characteristics of the chosen cooling technique.  To obtain a quantum circuit, the user must first instantiate one of the computational cooling modules with the constructed Cooling Unitary. The Dynamic Cooling module only requires the Cooling Unitary as input; the Suboptimal Dynamic Cooling module requires the user to provide a cluster size $n$ and number of steps $r$ in addition to the Cooling Unitary; the HBAC module requires a number of cooling rounds $r$ in addition to the Cooling Unitary; and finally, the Semi-Open AC module requires a number of cooling steps $r$, along with the cluster sizes $n_i$ that should be used in each step $i$.  If the cluster size is the same for every step, a single integer can be provided for the cluster size, otherwise, an array of clusters sizes should be provided where the $i^{th}$ entry corresponds to the cluster size for step $i$.   
After instantiation, a quantum circuit can be generated using the $getCircuit()$ method available within each cooling module.  This method relies the Gray code method described in Ref. \cite{bassman2024dynamic}.  For a given Cooling Unitary, Gray codes are generated for each permutation in the set.  A sub-circuit is created for each Gray code to implement the associated permutation using a series of multi-controlled-NOT (mcN) gates. Finally, all sub-circuits are concatenated together to generate the complete cooling circuit.  The circuit is output as a Qiskit circuit, which allows its easy execution on the IBM quantum simulators or quantum processors to assess the cooling performance and resilience to noise.  Other quantum backends usually provide functions to translate Qiskit circuits into their own type of native quantum circuits, expanding the set of quantum backends on which QuL circuits can be tested.

This approach to circuit generation is advantageous as circuits can be quickly created, even for large system sizes, which allows users to easily compare circuit sizes for various cooling techniques.  The drawback to this approach is that these circuits generally do not produce the shortest depth native-gate circuit implementation of the overarching unitary operator that defines the cooling procedure.  Note that the circuits output by QuL are comprised of multi-qubit gates which must be transpiled into a native, universal set of one- and two-qubit gates in order to run on a real quantum backend.  Therefore, before these circuits can be executed, one must transpile (i.e., convert) these complex gates into native gates (e.g., with IBM's Qiskit transpiler).  It has been observed that QuL circuits transpiled to native-gate circuits are generally larger in depth than native-gate circuits synthesized directly from the unitary operator defining the cooling procedure.  Therefore, when the qubit count is low, it may be advantageous to pass the unitary operator defining the cooling procedure directly to, for example, the Qiskit transpiler \cite{javadi2024quantum}, which will synthesize a circuit directly from the matrix.  While the circuit derived in this way may be smaller depth, the downside is that the for moderately large systems ($N \gtrsim 10$), the unitary operator has a very large memory footprint and the transpilation process requires large amounts of compute time.  
  
Every computational cooling module also provides methods to analyze key metrics of cooling performance.  These metrics include (i) the final temperature of the target qubit  (or equivalently the final probability of the target qubit being in the $1$ state), (ii) the energetic work cost associated with cooling, and (iii) the size of the quantum circuit.  Given an initial temperature for the system, each computational cooling module has its own method to compute the final temperature of the target qubit.  The work cost likewise requires the input of an initial temperature for the system, as well as the resonant frequency of the qubits, and computes the total amount of work expended during the cooling procedure.  Calculation of the work for the various methods are non-trivially different, as some methods involve cooling operators acting on qubits with different initial temperatures (e.g., in the second round of semi-open AC).  However, for a given cooling operator acting on qubits with a given set of initial temperatures, a value of work can be calculated according to Equation 9 Ref. \cite{bassman2024dynamic}.  The total work cost can thus be calculated by summing together the work contributions from each cooling operator included in the quantum circuit.  Finally, the size of the quantum circuit can be ascertained from the total number of multi-controlled-NOT gates present in the QuL circuit.  Using this count, users can quickly compare effective circuit complexities of various cooling procedures.  These counts can also give an indication of how resilient a cooling circuit will be to noise, as large gates counts imply larger error rates.

These analysis methods greatly simplify the task of comparing various cooling protocols (e.g., PPA, mirror protocol, minimum work protocol, etc.) and the different computational cooling techniques.  This will not only allow users to quickly identify the optimal cooling procedure for their given quantum backend, but also allow users to more easily investigate open questions in optimal cooling implementation.  It should be noted that the effects of noise on the various cooling procedures can also be tested by executing the derived quantum circuits on noisy quantum simulators, as will be shown in the illustrative examples below.

\section{Illustrative Example}
\label{sec:examples}
\subsection{Comparison of Final Temperatures}
\begin{figure}[h]
  \centering
  \begin{subfigure}[c]{0.45\textwidth}
        \includegraphics[width=\textwidth]{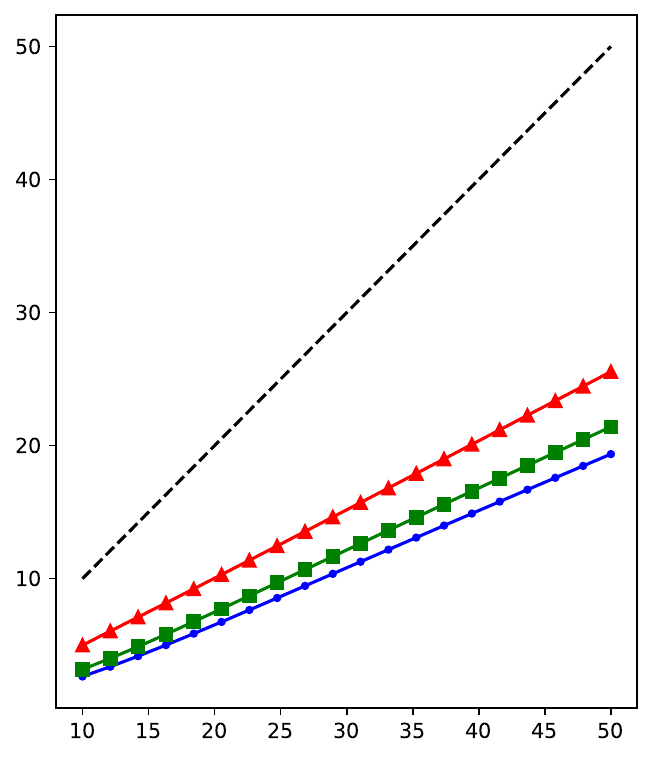}
        \put(-200,225){(a)}
        \put(-1,225){(b)}
        \put(-203,80){\rotatebox{90}{Final Temperature [mK]}}
        \put(-130,-5){Initial Temperature [mK]}
    \end{subfigure}
      \hspace{5mm}
  \begin{subfigure}[c]{0.4\textwidth}
    \includegraphics[width=\textwidth]{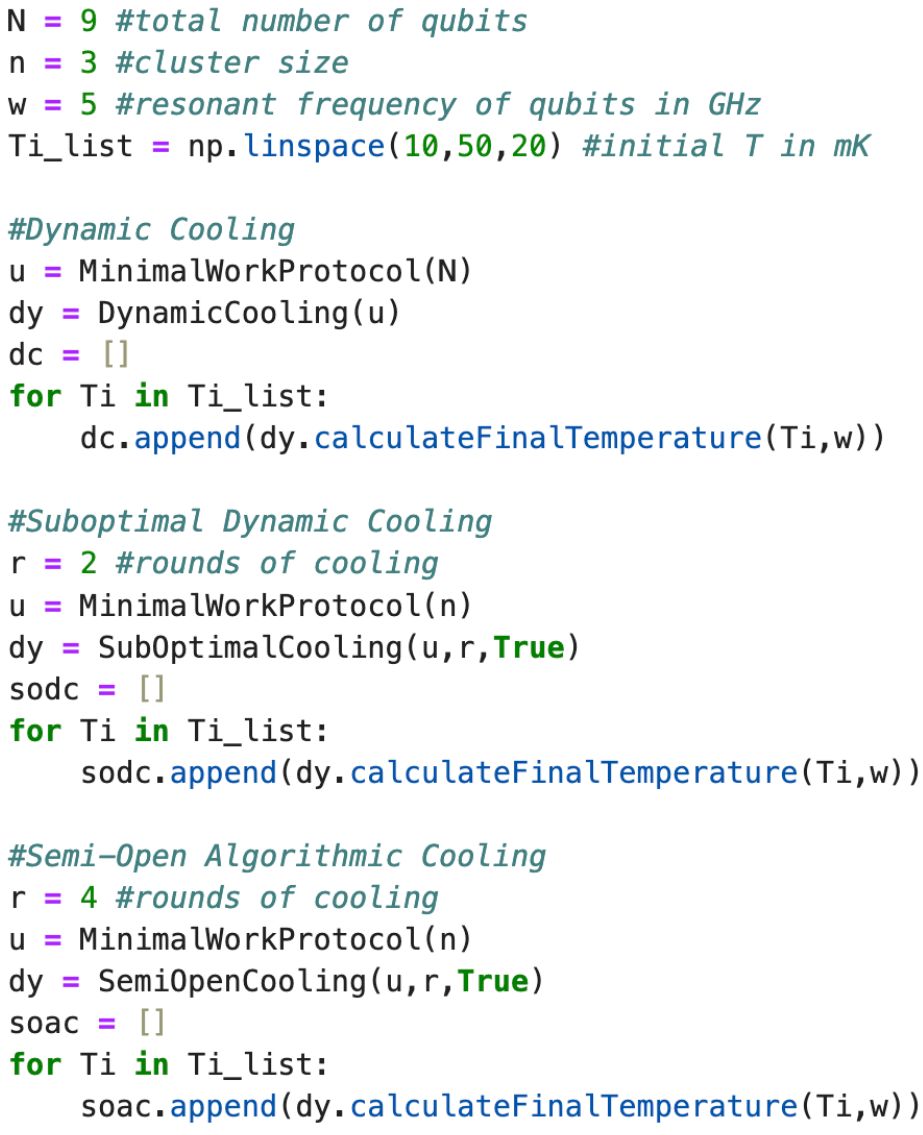}
    \end{subfigure}
  \caption{Comparison of final temperatures for various cooling protocols. (a) Final temperature of the target qubit after cooling via dynamic cooling (blue cirlces), suboptimal dynamic cooling (green squares), and semi-open algorithmic cooling (red triangles) for a range of initial temperatures.  The black dashed line plots final temperature equal to initial temperature and guides the eye to see the amount of cooling obtained by the various cooling methods. (b) Snippet of code using QuL to generate the data presented in (a).}
  \Description{Comparison of cooling methods}
  \label{fig:temp_comparison}
\end{figure}

In this example, we demonstrate how QuL can be used to quickly compare the ideal performance of various computational cooling protocols.  Here, we assume that we have access to a total of $N=9$ qubits and the minimal work protocol is chosen to generate the cooling unitaries.  Figure \ref{fig:temp_comparison}a shows the final temperature of the target qubit for a range of initial temperatures using dynamic cooling (blue circles), suboptimal dynamic cooling (green squares), and semi-open algorithmic cooling (red triangles).  The black dashed line plots the final temperature equal to the initial temperature, and guides the eye in seeing the amount of cooling performed.  The associated code for generating these curves is shown in Figure \ref{fig:temp_comparison}b.  For suboptimal dynamic cooling, we set the cluster size to $n=3$ and use $r=2$ rounds of cooling, which uses a total of $N=9$ qubits.  For semi-open algorithmic cooling, we use a cluster size of $n=3$ for each round of cooling, and by using $r=4$ rounds of semi-optimal AC, a total of $N=9$ qubits is again used for cooling.   While in each case the same total number of qubits is utilized, Figure \ref{fig:temp_comparison}a clearly shows that different computational cooling methods result in different final temperatures for the target qubit.  

We emphasize that these results represent cooling performance on an ideal quantum computer, i.e., one with no noise.  In practice, however, the systematic noise on quantum computers, as well as the connectivity of the qubits, will affect the performance of each method differently, since each method will generate a different quantum circuit.  In the next example, we show how QuL can be used to generate quantum circuits for various cooling protocols, which in turn can be used to gauge the effect of noise on different cooling protocols.

\subsection{Effect of noise}
\begin{figure}[h]
  \centering
  \begin{subfigure}[c]{0.5\textwidth}
        \includegraphics[width=\textwidth]{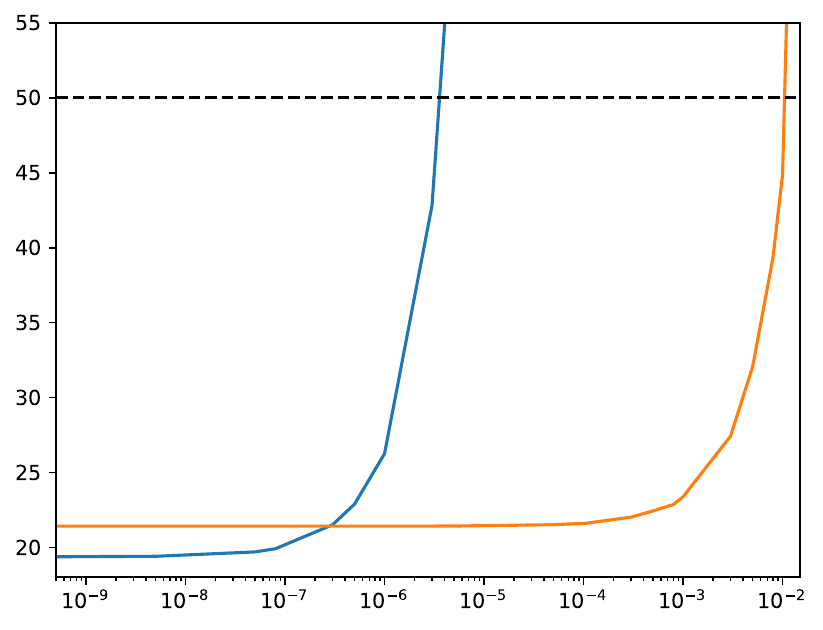}
        \put(-230,170){(a)}
        \put(-1,170){(b)}
        \put(-225,40){\rotatebox{90}{Final Temperature [mK]}}
        \put(-110,-1){$p$}
    \end{subfigure}
      \hspace{5mm}
  \begin{subfigure}[c]{0.35\textwidth}
    \includegraphics[width=\textwidth]{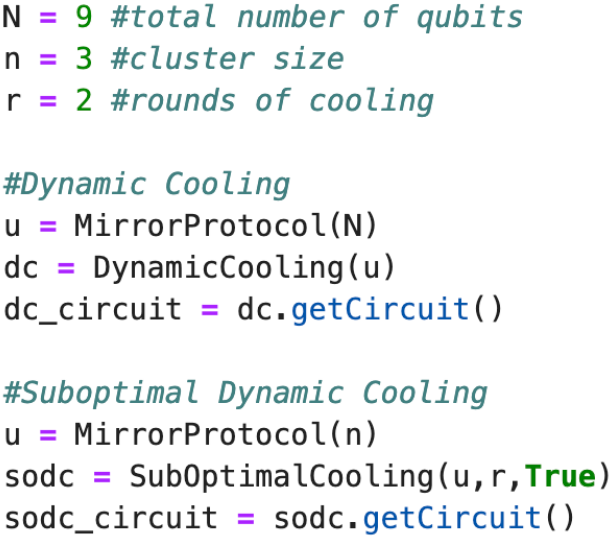}
    \end{subfigure}
  \caption{A comparison of the effect of noise for various cooling protocols.}
  \Description{Comparison of the effect of noise on different cooling methods}
  \label{fig:noise_comparison}
\end{figure}

In this example, we demonstrate how QuL can be used to generate quantum circuits for various computational cooling methods in order to better understand the effect of noise on the performance of different cooling protocols.  Figure \ref{fig:noise_comparison}a shows the final temperature of the target qubit versus a noise probability parameter $p$ for dynamic cooling with $N=9$ qubits (blue) and suboptimal dynamic cooling, also using a total of $N=9$ qubits, with a cluster size of $n=3$ and $r=2$ rounds of cooling (orange).  An initial temperature of 50 mK was used, indicated with the black dashed line.  The quantum circuits are generated with QuL using the mirror protocol following the code shown in Figure \ref{fig:noise_comparison}b.  They are executed on a simulated noisy quantum computer, using a noise model based on a depolarizing channel \cite{nielsen2002quantum}, which is tuned with noise parameter $p$, which effectively sets the probability of error. 

At very low levels of noise, Figure \ref{fig:noise_comparison}a shows that dynamic cooling achieves a lower final temperature than the suboptimal cooling scheme using the same total number of qubits.  However, as the level of noise is increased, the dynamic cooling protocol quickly loses superiority, since the quantum circuit for dynamic cooling is significantly larger than the one for suboptimal dynamic cooling.  Indeed, the suboptimal dynamic cooling protocol remains nearly unaffected by noise until large error probabilities are reached.  This implies that on noisy quantum hardware, for a given number of qubits, suboptimal dynamic cooling provides significantly increased cooling capability over the standard dynamic cooling method, even though in principle it cannot reach as low of a final temperature.


\section{Conclusion}\label{sec:conclusion}
QuL is an extensible, open-source programming library for generating quantum circuits for computational cooling, as well as for analyzing these circuits across various metrics. By simply providing the desired number of qubits and selecting a built-in protocol, a user automatically produce a variety of cooling circuits and easily assess their performance across various metrics in order to determine optimal computational cooling protocols given various resource constraints or particular quantum backends.  QuL is also flexible enough to allow more advanced users to implement and test their own cooling protocols.  The full code, along with a tutorial, can be found on GitHub \cite{github}.  

The current release comprises four main computational cooling methods, namely, dynamic cooling, suboptimal dynamic cooling, heat-bath algorithmic cooling, and semi-open algorithmic cooling.  Each method can be derived from one of three built-in cooling protocols, namely, the partner-pairing algorithm, the minimal work protocol, and the mirror protocol.  QuL was intentionally designed in this modular fashion to enable the seamless integration of novel cooling protocols and cooling methods, as they are developed. Indeed, we anticipate researchers will be able to leverage QuL to derive novel cooling techniques.

\begin{acks}
LBO gratefully acknowledges funding from the European Union’s Horizon 2020 research and innovation program under the Marie Skłodowska-Curie grant agreement No 101063316.  This research used resources of the National Energy Research Scientific Computing Center (NERSC), a Department of Energy Office of Science User Facility using NERSC award DDR-ERCAP29585.  This research also used resources of the Oak Ridge Leadership Computing Facility, which is a DOE Office of Science User Facility supported under Contract DE-AC05-00OR22725. We acknowledge the use of IBM Quantum services for this work. 
\end{acks}


\bibliographystyle{ACM-Reference-Format}
\bibliography{references}

\end{document}